\documentclass[a4paper,12pt]{article}
\usepackage[utf8]{inputenc}
\usepackage[margin=1.25in]{geometry}
\usepackage{mathptmx}

\usepackage{endnotes}
\let\footnote=\endnote

\usepackage{natbib}
\usepackage{amsmath,amssymb}
\usepackage{graphicx}
\usepackage{color}
\usepackage{natbib}

\usepackage{setspace}

\usepackage{tcolorbox}

\usepackage{siunitx}
\usepackage{booktabs}
\usepackage{tabularx}
\usepackage{tabulary}
\usepackage{rotating}

\usepackage{multirow}
\usepackage{subfigure}
\usepackage{color, colortbl}
\usepackage{url}

\title{Statistical Power in Longitudinal Network Studies}
\date{}

\usepackage{authblk}
\author[1]{Christoph Stadtfeld%
\footnote{Corresponding author: Christoph Stadtfeld,
ETH Z\"urich,
Chair of Social Networks,
Department of Humanities, Social and Political Sciences,
Clausiusstrasse 50,
8092 Zürich, Switzerland, 
phone: +41 44 632 07 93,
e-mail: c.stadtfeld@ethz.ch}
}
\author[2,3]{Tom A. B. Snijders}
\author[2,4]{Christian Steglich}
\author[2]{Marijtje van Duijn}
\affil[1]{\small{Chair of Social Networks\\
		ETH Z\"{u}rich}, Switzerland}
\affil[2]{\small{Department of Sociology, University of Groningen, Netherlands}}
\affil[3]{\small{Nuffield College, University of Oxford, UK}}
\affil[4]{\small{Link\"oping University, Institute for Analytical Sociology, Sweden}}

\graphicspath{ {} {img/} }

\begin{document}

\maketitle

\begin{tcolorbox}
Uncorrected preprint. Please cite as:\\
 Stadtfeld, Christoph, Tom A. B. Snijders, Christian Steglich \& Marijtje van Duijn. Forthcoming. ``Statistical Power in Longitudinal Network Studies''. \emph{Sociological Methods and Research}
\end{tcolorbox}
  
\begin{abstract}
Longitudinal social network studies can easily suffer from insufficient statistical power.
Studies that simultaneously investigate change of network ties and change of nodal attributes (selection and influence studies) are particularly at risk because the number of nodal observations is typically much lower than the number of observed tie variables.
This paper presents a simulation-based procedure to evaluate statistical power of longitudinal social network studies in which stochastic actor-oriented models (SAOMs) are to be applied.
Two detailed case studies illustrate how statistical power is strongly affected by network size, number of data collection waves, effect sizes, missing data, and participant turnover.
These issues should thus be explored in the design phase of longitudinal social network studies.
\end{abstract}


\newpage
\doublespacing

\section{Introduction}
Longitudinal social network studies are costly and time-consuming both for researchers and participants. A lack of statistical evidence for a hypothesis should thus not originate from a study design that was ``just too small'' and, therefore, has insufficient statistical power \citep{Cohen1977}. 

The introduction of Stochastic Actor-Oriented Models for the simultaneous investigation of network and behavior changes \citep[SAOMs,][]{Snijders2010, Steglich2010} enabled a large number of publications that empirically study selection processes (changes in social relations in response to individual attributes) and influence processes (changes in individual attributes in response to social relations). 
SAOMs are typically applied to network panel data (a set of interconnected individuals surveyed in multiple data collection waves) and evaluate dynamic tendencies of individuals to change (add or drop) network ties and to change (increase or decrease) some type of behavior or individual attribute.
\citet{Veenstra2013} review a number of selection and influence studies on adolescent peer relations\footnote{A nearly complete list of SAOM-related publications is available at \citet{Snijders2016}.} and report mixed evidence regarding the prevalence of selection and influence mechanisms in adolescent behaviors, by finding significant effects in some and non-significant effects in other studies. 
It is possible that some of the studies were underpowered, however, until now there has been no method to perform power analyses for study designs in longitudinal network research.

Indeed, statistical power might be particularly hard to achieve in social networks studies that do not only consider network change (e.g., friendship relations) but also change in individual attributes (e.g., the level of delinquency).
At each data wave, $N$ nodes are connected through multiple network ties. When $\overline{k}$ is the average degree (it is typically larger than one in meaningful network studies) this results in $N \cdot \overline{k}$ tie observations and a high number of observations of non-existing ties ($N \cdot (N-1)$ tie variables in total). In comparison, only N nodal attributes are observed per data wave\footnote{This observational asymmetry was discussed by \citet{Krivitsky2015}.}. This implies generally less information available in the estimation of behavior change mechanisms and in consequence also lower power to detect these mechanisms.

This paper introduces a procedure for power analyses of longitudinal network studies that make use of SAOMs in the empirical analysis. 
It further aims at providing some guidelines for researchers who are designing new studies and raising awareness about critical issues such as missing data and participant turnover.

In classic power studies \citep[see, for example,][]{Cohen1977} power depends on three parameters: the \emph{significance level}, \emph{sample size} and \emph{effect size}.
Recall that the significance level $\alpha$ is known as type I error, the probability to (incorrectly) reject the null hypothesis when it is true. Power is defined as the probability to (correctly) reject the null hypothesis when the alternative hypothesis is true, also known as 1-$\beta$ or 1 - type II error, where type II error is defined as the probability to (incorrectly) not reject the null hypothesis when it is not true. To compute the power, the alternative hypothesis needs to be specified. The effect size is a measure for the difference or distance between the null and the alternative hypotheses.

Although power analyses have been developed for study designs with simple random or clustered data, social network data are characterized by a more complex dependence structure requiring a more involved method to estimate power. 
While in SAOMs parameter estimates can be tested at the customary 0.05 significance level (using approximate t-tests), the definition of sample size and the effect size require some more elaboration. 

The ``sample size'' in dynamic social network studies is affected by a number of aspects that we refer to as the \emph{study design}. Larger studies with many individuals, joint analysis of multiple networks, and several data collection waves will exhibit more statistical power than small-scale studies. But also design decisions about the granularity of a behavioral scale or a maximum number of nominations in a questionnaire may affect the statistical power.
``Sample size’’ is a concept originating from statistical models constructed of independent observations, and is not directly applicable to network studies. \citet{Krivitsky2015} discuss the question what sample size could mean for network studies, and limit their interpretation of effective sample size to ''the scaling of the asymptotic variances of maximum likelihood estimates in a network model`` (op.\ cit., p.\ 186). A summary of their main conclusion is that this will be of the order of $N$ for sparse and of $N^2$ for non-sparse network data. This is not directly helpful for stochastic actor-oriented models because of the dynamic nature of the data under study. 
However, the authors' experience 
suggests that the scaling of the amount of information, or the inverse of variances of parameter estimates, for SAOMs for sparse network data will very approximately be proportional to $N \times {\bar k} \times (M-1) $, where $N$ is the number of nodes, 
${\bar k}$ the average degree, and $M$ the number of waves. 
This approximation applies only to the network parameters, not to the behavior parameters. 
The presumed dependence on the average degree $\bar k$ is tentative, and should be further investigated; there will be a quite strong dependence on whether $\bar k$ is invariant with respect to network size (e.g., as in case of resource constrained networks like friendship networks), on other features of the network structure and the distribution of the behavior, which may in some cases be stronger than the dependence on the average degree.

The ``effect size'' (usually, a difference in means or a strength of association) is also somewhat more involved in dynamic network studies where a high number of social mechanisms simultaneously operate that confound, interact with, or amplify one another.
For SAOMs, standardized effect sizes have not yet been developed, and therefore the values of the model parameters must be used as effect size measures. 
The parameters should be informed by empirical SAOM results. It should be taken into account that parameter estimates are (as in any statistical model) depending on the scaling of variables or the size and distribution of opportunity sets, thus a similar empirical setting should be chosen.
The chosen parameters will matter for the power of a social mechanism. For example, strong social influence mechanisms that operate almost deterministically will be easier to discover than subtle mechanisms.
Social mechanisms that interact with the behavioral outcomes of theoretical interest (e.g., homophily on a correlated variable), or mechanisms that amplify the level of observed similarity of connected nodes \citep[e.g., transitivity, see][]{Stadtfeld2015a} will potentially reduce the statistical power of the mechanism within the proposed model and should thus also be considered.
The statistical power is further affected by interfering mechanisms such as participant turnover rates and non-response.

Researchers typically have various options on how to define a study design (conditional on their theories and research questions), while facing uncertainty about the social mechanisms that operate in their sample.
The distinction between the two dimensions is not necessarily sharp. For example, researchers may be able to reduce non-response (an interfering data collection mechanism that reduces the ``sample size'') through changes in their study design by, for example, facilitating participation through online access, simplifying questionnaires, or incentivizing participation. 
Yet we think that the distinction between study design decisions and uncertainty about social mechanisms is conceptually helpful as it is in line with the traditional notion of power studies that are concerned with sample size (a study design decision) and effect sizes (which refer to assumptions about the strength of social mechanisms of interest).

The proposed procedure for the evaluation of statistical power in longitudinal network studies consists of six steps and is introduced in section \ref{sec:simulation_process}. 
The procedure makes use of the R package NetSim \citep{Stadtfeld2015} to simulate social network data, and of the R package RSiena \citep{Ripley2016} to simulate and estimate SAOMs. 
To illustrate the six-step procedure, we discuss two empirically inspired research settings in sections \ref{sec:local_communities} and \ref{sec:smoking} that are in line with what we perceive as ``typical'' empirical selection and influence studies. 
The first research setting in section \ref{sec:local_communities} examines how the number of data collection waves and the delineation of a network affect the statistical power. This research setting relates to exploring alternative \emph{research designs} (the ``sample size'').
The second research setting in section \ref{sec:smoking} discusses statistical power of selection and influence effects in an empirical setting with social networks collected in multiple schools. In particular, we investigate to what extent statistical power is influenced by homophily and social influence effect sizes, by respondent data that are missing completely at random \citep{Huisman2008, Haye2017}, and by turnover of students between data collection waves \citep{Huisman2003}. This research setting relates to exploring a space of varying \emph{social mechanisms} (the ``effect sizes'').
The two exemplary research settings illustrate how power analyses can be applied in practice and address specific issues that researchers should be concerned about. 
However, they do not aim at exploring the relation between assumptions about social mechanisms and possible research designs in full depth as those will be highly context dependent. Our findings indicate that considering issues like network size, number of data collection waves, participant turnover, missing data, and effect sizes are of critical importance in the design phase of longitudinal network studies. 
Section \ref{sec:conclusions} discusses the potential impact of this paper on the design of future longitudinal social network studies.

\section{A procedure for the estimation of statistical power}
\label{sec:simulation_process}

\begin{figure}[htbp]
  \includegraphics[width = \textwidth]{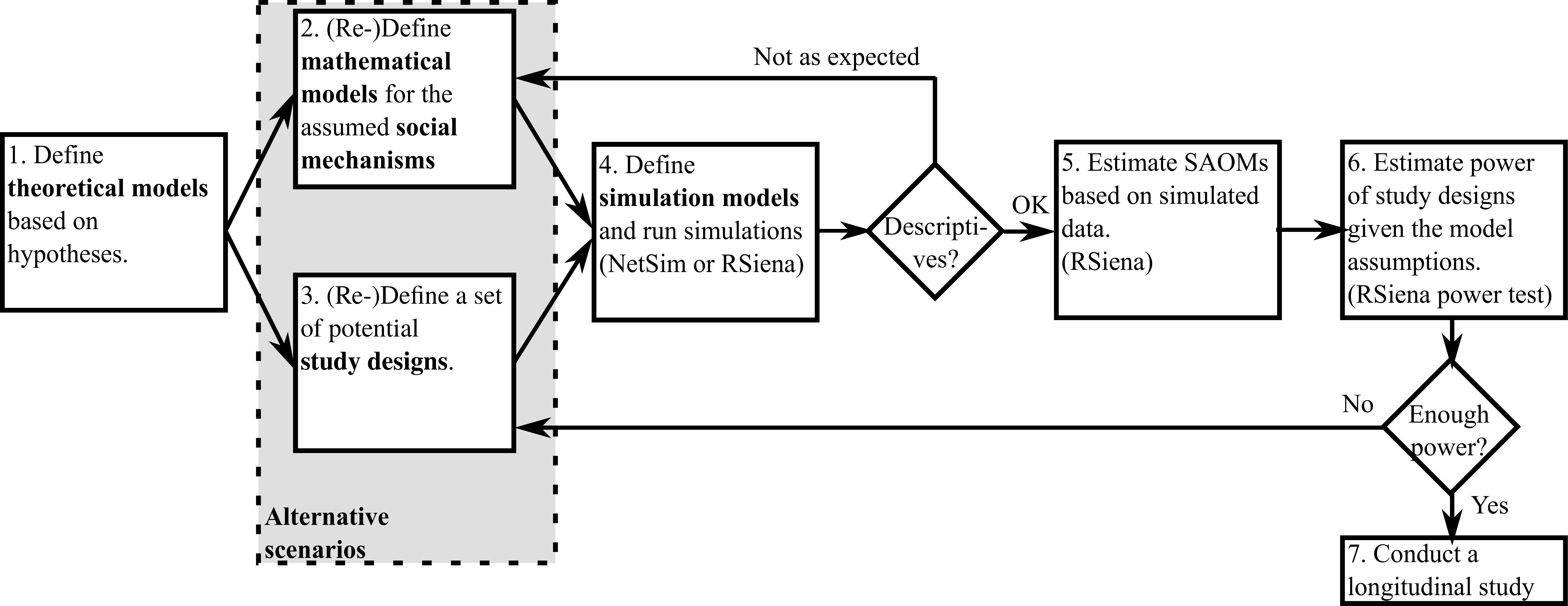}
\caption{Overview of the procedure for the estimation of statistical power in longitudinal social network studies.}
\label{fig:process}
\end{figure}

The proposed procedure evaluates a range of alternative scenarios that 
vary in research designs and express uncertainty about the prevalence and magnitude of various social mechanisms. The procedure is sketched in Figure \ref{fig:process} and consists of six major steps.

\begin{enumerate}

\item Each longitudinal social network study starts with the formulation of hypotheses on social mechanisms. Typical hypotheses relate to homophily processes in the network formation \citep{McPherson2001} and social influence processes on the attribute level \citep{Friedkin1998}. 
However, many other research questions in the domain of social networks can be considered. Those can relate to network change processes, such as reciprocity, transitivity, or popularity mechanisms \citep{Kadushin2012}, or to attribute change processes. 

\item[$\rightarrow$]The following two steps span a space of \emph{alternative scenarios} for which statistical power analyses can be performed. 

\item The \emph{social mechanisms} identified in step~1 are translated into formal \emph{mathemati\-cal models}. The class of stochastic actor-oriented models (SAOMs) is a good starting point as it allows the combination of several network- and attribute-related social mechanisms \citep{Snijders2010, Steglich2010, Snijders2015}. 
But also other mathematical frameworks could be applied, for example, 
tie-based Markov models that generate Exponential Random Graph distributions \citep[][ch.12]{Block2016, Lusher2013}, micro-models proposed for network event models \citep{Butts2008, Stadtfeld2017}, or Hierarchical Latent Space Models \citep{Sweet2013, Sweet2016}.
It is possible that some aspects of the theoretical model cannot be expressed with SAOMs, for example, processes that lead to specific types of missing data or cause individuals to join and leave the population. 
Processes of that kind can be formalized outside of the SAOM framework as illustrated in section~\ref{sec:smoking}.
Good a-priori expectations about social mechanisms and their effect sizes are difficult, especially in view of the high interdependence between model parameters. 
As a pragmatic starting point, ranges of parameters found in prior empirical studies may be chosen as effect sizes whereby research on SAOM parameter interpretation (as discussed in \citet[section 3.4]{Snijders2010} and \citet[chapter 13]{Ripley2016}) should be taken into account.
The research setting in section~\ref{sec:smoking} focuses on this step~2.

\item Potential \emph{study designs} are defined to address the hypotheses formulated in step~1. A first ad-hoc attempt may build on designs of previous research studies. Typical decisions in this step are defining the number of individuals in the study (i.e., number of networks or network boundaries), prolonging the study by increasing the number of waves of data collection, intensifying the study by reducing the time spans between subsequent waves, changing the granularity of a behavioral scale, or deciding whether the number of nominations in a network questionnaire should be restricted.
Research design decisions are naturally constrained by the theoretical framework and the empirical setting of a study.
The research setting in section~\ref{sec:local_communities} focuses on this step~3.

\item Simulation models are defined for a reasonable subset of the alternative scenarios described by steps~2 and~3. Additional assumptions may be necessary. These may relate to starting distributions of individual attributes or network structures at the beginning of a data collection (such decisions could be based on theoretical expectations or prior empirical work). For each simulation model a number of simulations is run (e.g., 200). Descriptives of the simulated networks and individual attributes should be checked at the end the simulations to determine if the simulations generate unexpected or unrealistic outcomes. 
One could, for example, check whether clustering or degree distributions are in a range that is found in comparable studies and is in line with theoretical expectations. 
This can be done in RSiena using the sienaGOF (``Goodness-of-fit'') function, which gives the distribution of statistics; the comparison with a true observed value is not relevant for this use of sienaGOF.
If descriptives of the simulated networks are unreasonable, the mathematical models from step~2 should be improved. In this paper, we simulate data with the R package NetSim \citep{Stadtfeld2015} and the RSiena package \citep{Ripley2016}. RSiena can be used to simulate SAOM processes. In case other social mechanisms are to be simulated (for example, processes that explain composition change or missing data), more general packages such as NetSim can be applied. 
Previous papers in which RSiena was applied in simulation studies are \citet{Snijders2015} and \citet{Prell2016}.
Example simulation scripts with RSiena and NetSim are 
published online\footnote{Scripts and supplementary material will be published online with publication of the paper.}.

\item The simulated data sets (say, 200 per simulation model) are used as data input for an estimation with the RSiena software. 
Stochastic actor-oriented models are specified according to the theoretical models in step~1. This step of re-estimating models may take a considerable amount of computation time as the number of simulation models is relatively large and the simulation-based estimation of parameters of the SIENA method is time-consuming. However, by using parallel computing the effective computing time can be largely reduced.
 
\item For each SAOM fit to the simulated data sets, the percentage of cases is calculated in which significant parameters were estimated in the re-estimation step~5. 
The statistical power evaluation will firstly focus on social mechanisms about which hypotheses have been formulated, even though the procedure can be
valuable to explore how a study design is likely to impact the interpretation of other effects in the model.
The significance can, for example, be tested at a $\alpha=0.05$ significance level. 
A more efficient estimator could be given by estimating the mean and standard deviation of the parameter estimate or the mean of the t-ratios (with assumed variance 1) and estimate power from 
there\footnote{
The tests used for the SAOM are approximate $t$-test based on the ratio
$t = \hat{\beta}/\mbox{S.E.}(\hat{\beta})$.
For such tests we have the well-known formula \citep[see][p.178]{Snijders2011}
\begin{equation}
    \frac{\mbox{parameter}}{\mbox{standard error}} \approx
       z_{1-\alpha} + z_{1-\beta} =  z_{1-\alpha} - z_{\beta}, \label{equ:power}
\end{equation}
where $\alpha$ is the significance level and $1-\beta$ the power of the test,
while $z_{1-\alpha},\, z_{1-\beta}$, and $z_{\beta}$
are the values from the standard normal distribution
associated with the cumulative probability values indicated.
This formula can be used for a more efficient estimator from the simulation results.
In equation~\ref{equ:power}  we insert the mean parameter estimate as the parameter,
and the standard deviation of the parameter estimates as the standard error,
and given the intended $\alpha$ we can calculate the power  $1-\beta$.}.
The percentage of (correctly) rejected null hypotheses (of no effect) is an estimate of the statistical power of the study design. 
If several study designs seem to provide satisfactory power, then the least costly can be chosen and the longitudinal study can be conducted. If the power in all study designs is too low, then changes should be considered. This corresponds to updating the study designs in step~3. 
\end{enumerate}

%

\section{Research setting 1: Opinion dynamics in four local communities}
\label{sec:local_communities}
The first research setting discusses a (fictitious) study design in which the dynamics of friendship and opinion formation (negative -- neutral -- positive) in four local communities are observed. The communities are geo-spatially close to one another so that interpersonal ties may occur between them, however, ties within communities are more likely.
We sketch a research study in which the friendship network and opinion dynamics of 120 individuals are of interest. The key hypotheses are that both homophily and influence processes with regards to opinions are prevalent. The design decisions take the network boundaries and the number of waves of data collection into account.
To investigate the statistical power of different study designs, we follow the six-step procedure introduced in section~\ref{sec:simulation_process}.

\subsection{Hypotheses and assumptions}
In this study we are interested in two hypotheses, namely whether changes in opinions are explained by the opinions of friends (social influence) and whether individuals choose their friends based on opinion similarity (homophily). Several additional dynamic assumptions are made. 
These are chosen with the purpose to demonstrate how specific processes of social influence can be tested within a SAOM framework.
First, we assume that individuals have a slight tendency for polarization. In the absence of social influence effects (e.g., when individuals are not connected to others), individuals are expected to have a slightly higher propensity to develop extreme opinions (negative or positive instead of neutral). Second, we assume a friendship network formation that is partly driven by preferences for reciprocity, geo-spatial proximity (propinquity) and by preference for transitive structures. Third, personal networks of individuals are assumed to change faster than their opinions. Furthermore, we start with some straightforward assumptions about how the friendship network and the distribution of opinions look like at the beginning of the study. 

\subsection{Mathematical formulation}
The hypotheses and the additional assumptions are formalized as a stochastic actor-oriented model (SAOM). Based on the parameters of ``typical'' empirical SIENA models\footnote{The model is inspired by parameters and model specifications found in empirical studies. Overviews are provided by \citet{Snijders2010} and \citet{Veenstra2013}. For example, transitivity parameters of about 0.2 and reciprocity parameters of about 2 have been reported in a variety of studies. The SIENA webpage \citep{Snijders2016} further lists the majority of papers that apply SAOMs.}, we formalize the exemplary model with the specification shown in Table~\ref{tab:specification1}. 
Parameters were further adjusted so that when simulated, the model would not be ``degenerate'' in a sense that it is unlikely to generate networks that have a density close to one or zero.
The question how to translate hypotheses into SAOM parameters is nontrivial -- empirical findings of studies in related empirical and theoretical contexts can provide reasonable starting values \citep[for an overview we refer to the SIENA website,][]{Snijders2016}. 
The opinion variable is assumed to be measured on a three point scale from one to three.

\newcommand{\code}{\texttt} 
\begin{table}[htb]
\renewcommand{\arraystretch}{1.3}
\begin{center}
\begin{tabular}{rllr}
   & \textbf{Mechanism} & \textbf{SIENA effect name} &\textbf{Parameter}\\
\hline
   Friendship & Change & \code{rate} & ~3.0 \\
   &Density & \code{density} & -2.0\\
   &Reciprocity & \code{recip} & ~2.0\\
   &Transitivity & \code{transTrip} & 0.2 \\
   &Cyclic closure & \code{cycle3} & -0.1 \\
   &Propinquity (Distance) & \code{X} & -2.5\\
   & \emph{Homophily} (Opinion) & \code{simX} & 1.5 \\
   Opinion & Change & \code{rate} & 0.6 \\
   & Center & \code{linear} & -0.8\\
   & Dispersion & \code{quad} & 0.2 \\
  &  \emph{Influence} & \code{totSim} & 0.8
\end{tabular}
\end{center}
\caption{Specification of a stochastic actor-oriented model that expresses assumptions about the social mechanisms at play (step~2) in the first research setting. The focal mechanisms are emphasized.}
\label{tab:specification1}
\end{table}

\subsection{Research designs}
We explore two types of design decisions. The first design decision is about the friendship network delineation: Should data be collected in one, two or all four local communities (N = 30, 60 or 120)? 
We assume that the social mechanisms sketched in the previous section govern the social processes in the whole sample of 120 individuals (four communities), but discuss study designs that collect data just within one or two sub communities (30 or 60 individuals).
The second design decision is concerned with the number of data collection waves. In this example, we consider collecting two waves, three waves or five waves of data. 
By adding more data collection waves, the duration of the study is extended: data collection waves are not added in-between two waves but increase the duration of the data collection period by factor two or four. The time between two sub-sequent data collections is the same across all study designs.

\subsection{Simulation models}
We generate five simulation models based on the mathematical formulation and a subset of the space of potential study designs. 
The five simulation models relate to five study designs and are sketched in Table \ref{tab:simulations1}. 
From each simulation model 200 data sets are generated with the software package NetSim \citep{Stadtfeld2015}\footnote{In this example in which the mathematical model is completely in line with the SAOM framework, the RSiena software could have been used for simulations as well}. The simulation is always run on the complete data set of 120 nodes and only then sub samples (regarding number of waves and network delineation) are drawn.

\begin{table}[htb]
\renewcommand{\arraystretch}{1.3}
\begin{center}
\begin{tabular}{lccc}
   &\textbf{four communities} & \textbf{two communities} &\textbf{one community}\\
   &\textbf{(N = 120)} & \textbf{(N = 60)} &\textbf{(N = 30)}\\
   \textbf{5 waves} & X& & X\\
   \textbf{3 waves} & &X & \\
   \textbf{2 waves} &X & &X \\
\end{tabular}
\end{center}
\caption{Five out of nine possible simulation models are chosen.}
\label{tab:simulations1}
\end{table}

\begin{figure}[htbp]
   \subfigure[Wave 1]{
    \includegraphics[width=0.55\textwidth]{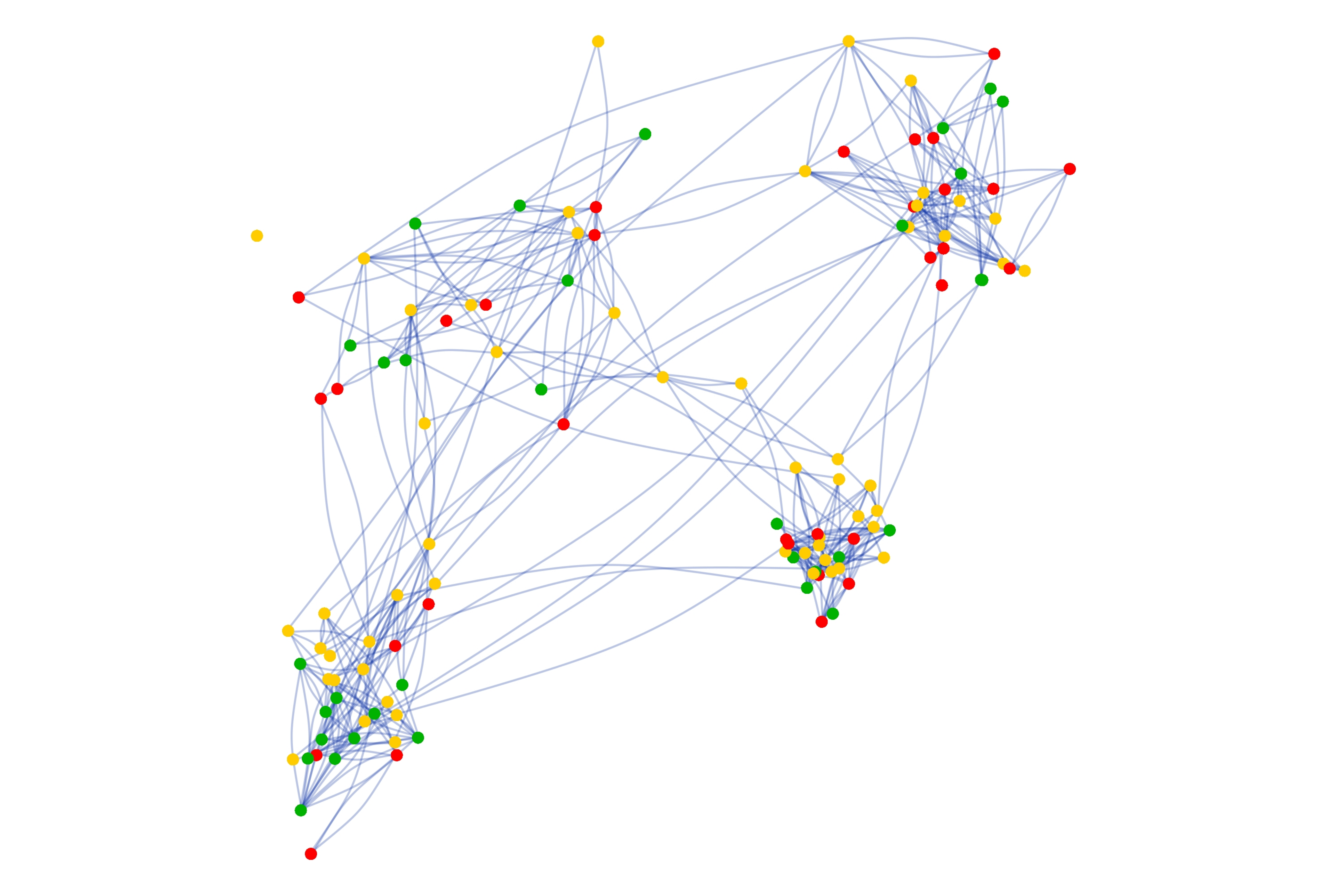}
  \label{fig:wave1}
  }
   \subfigure[Wave 2]{
    \includegraphics[width=0.55\textwidth]{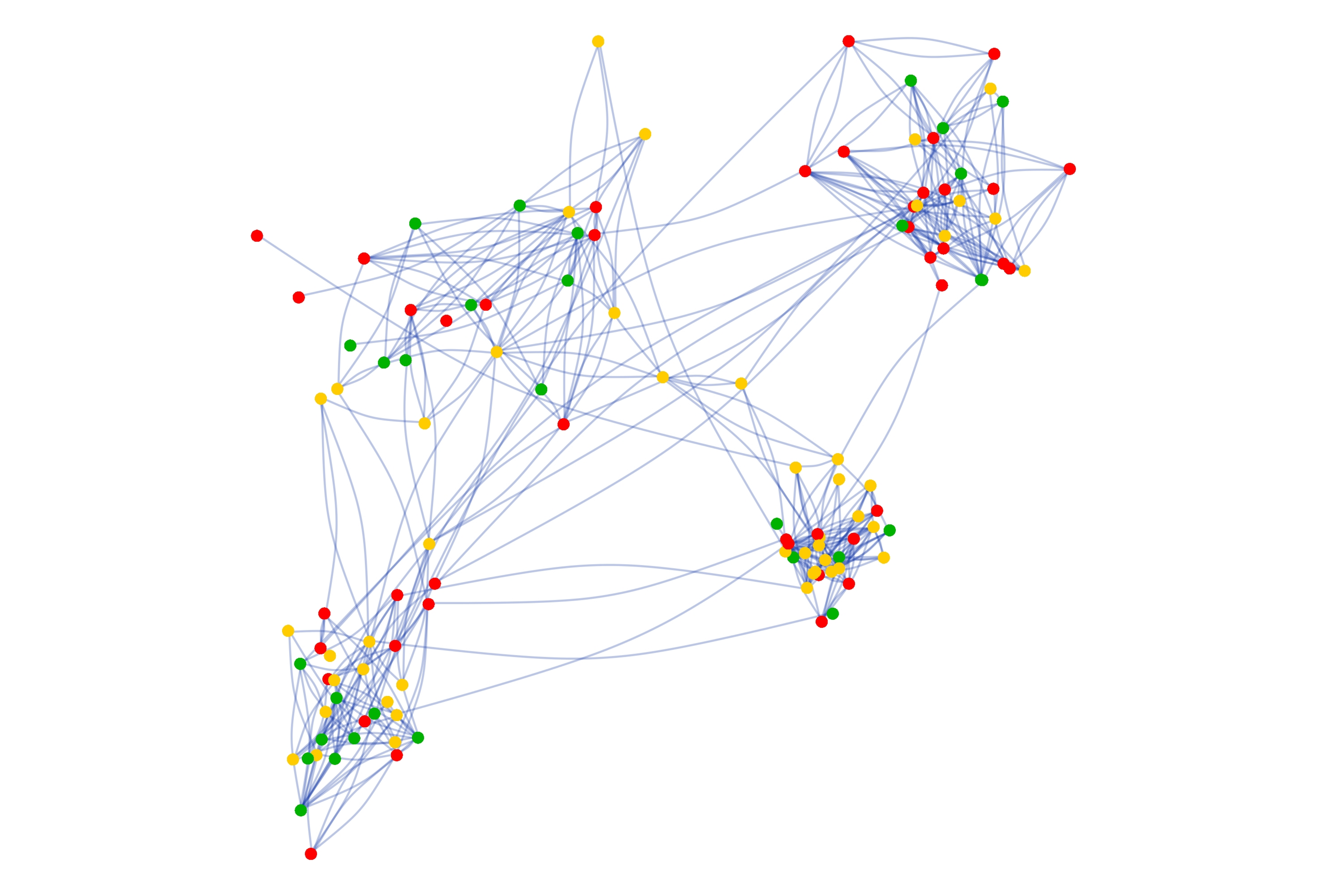}
  \label{fig:wave2}
  }
   \subfigure[Wave 3]{
    \includegraphics[width=0.55\textwidth]{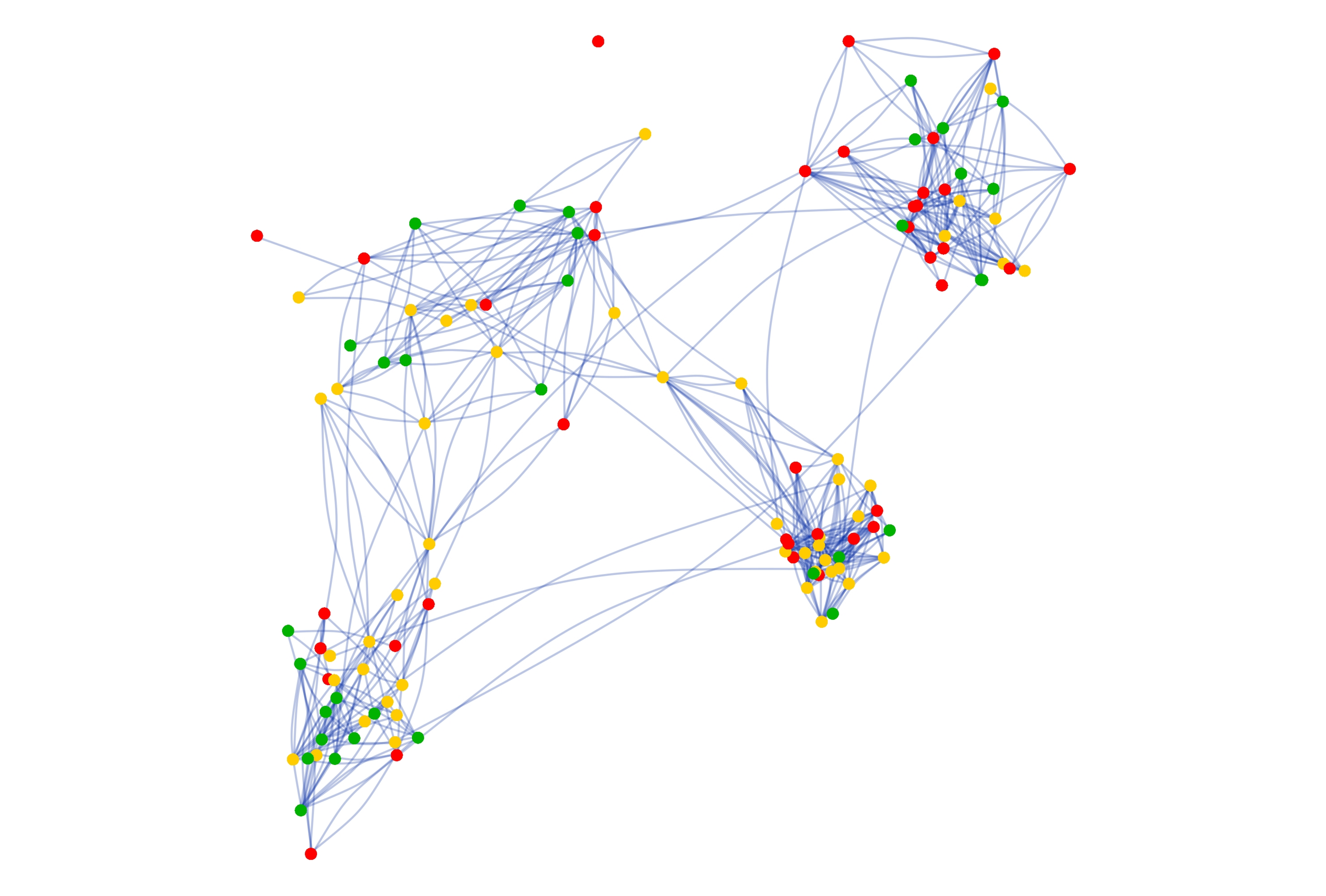}
  \label{fig:wave3}
  }
   \subfigure[Wave 4]{
    \includegraphics[width=0.55\textwidth]{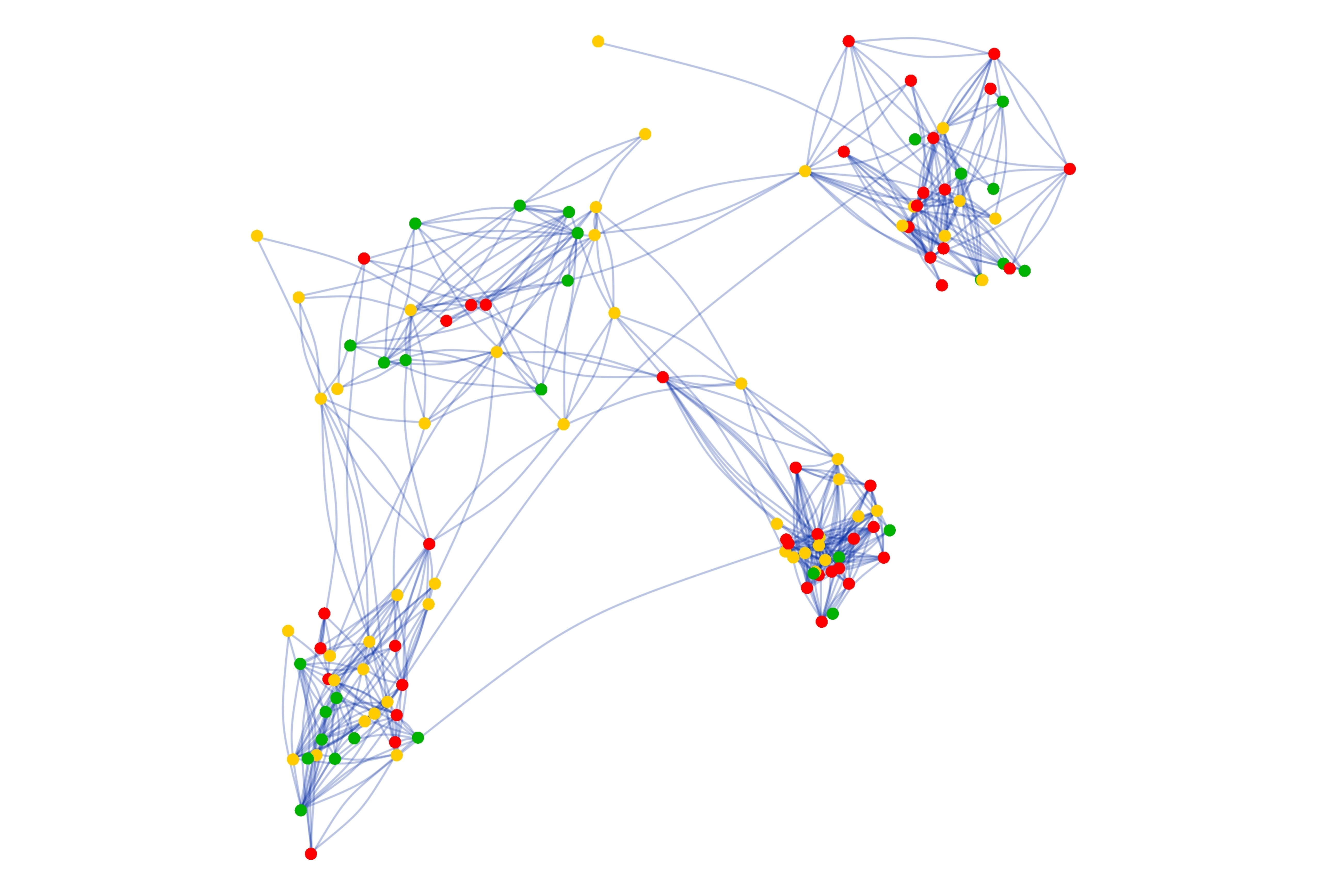}
  \label{fig:wave4}
  }
\caption{Four waves of data generated by the simulation process in one simulation run. Both the friendship network and the attributes (indicated by color codes) change over time following the model specified in Table~\ref{tab:specification1}. All four local communities are shown. The network layout corresponds to the geo-spatial distribution of individuals in the study.}
  \label{fig:waves}
\end{figure}

Each simulation is based on an initial equal distribution of opinions and an initial friendship network. The starting network is simulated from an empty network with the stochastic actor-oriented model shown in Table~\ref{tab:specification1}, except for the homophily and influence effects. After this initial process that is run until the network has a stable density, 
individual attributes are randomly assigned to actors in order to achieve an initial observation in which network position and individual attributes are uncorrelated. This relates to an assumption made in this study that social effects on opinion formation only start playing out after the initial data collection. Figure \ref{fig:waves} shows four networks that were extracted from one simulation run. Actors are positioned in a two-dimensional space; the distance between actors affects the propensity to form network ties. Locations are randomly drawn from four two-dimensional normal distributions with different means and variances.
Checks of network densities and degree distributions reveal that the simulated networks are 
reasonable from a descriptive point of view. In particular, the simulation model is not degenerative in a sense that it would produce graphs with a density close to one in the long run.
Therefore, we proceed with step 5 of the procedure. A visualization of a related dynamic four-community simulation can be found in the online appendix\footnote{A simulation video based on the NetSim package is published with this paper.}. It demonstrates the non-degeneracy of the specified model.

\subsection{Estimation with RSiena}
After the simulations, the generated data are fitted to a stochastic actor-oriented model using the RSiena software. This model is specified with exactly the same parameters that were used in the mathematical model (see Table \ref{tab:specification1}). The simulation phase generated 1000 result sets ($5 \times 200$) that include parameter estimates and standard errors. This process takes a significant amount of time (about one day on a standard personal computer) but can be accelerated by making use of parallel computing. All 1000 simulations and subsequent estimations with RSiena are independent and can thus be processed in parallel. This means that step~5 can be processed in much less than one hour in this case study.

\subsection{Evaluating the power}

For each simulation model, the power of the parameters is evaluated. As an example, the results of the scenario with two local communities (N = 60) and three waves of data collection are shown in Table~\ref{tab:results1}. It includes the effect names, the simulation model parameters (see Table \ref{tab:specification1}), the mean estimated parameters of 200 simulated data sets, their standard deviation and the power of the effects in this particular study design. The power column indicates the percentage of simulated data sets for which a parameter was re-estimated significantly with a p-value smaller than $0.05$.

\begin{table}[htb]
\begin{center}
\begin{tabular}{rlrrrr}
&\textbf{Effects} & \textbf{Sim. param.} & \textbf{Avg. est} & \textbf{St. dev} & \textbf{Power (\%)}\\ 
  \hline
  Friendship & Change & 3.0 & 2.49 & 0.32 & \\ 
  &Density & -2.0 & -3.17 & 0.19 & 100.0 \\ 
  &Reciprocity & 2.0 & 2.08 & 0.18 & 100.0 \\ 
  &Transitivity & 0.2 & 0.27 & 0.09 & 85.0 \\ 
  &Cyclic closure & -0.1 & -0.19 & 0.16 & 17.5 \\ 
  &Propinquity (Distance) & -2.5 & -1.18 & 0.16 & 100.0 \\ 
  \rowcolor[gray]{.9} &Homophily (Opinion) & 1.5 & 1.55 & 0.40 & 99.5 \\ 
  Opinion&Change & 0.6 & 0.58 & 0.21 & \\ 
  &Center (linear) & -0.8 & -0.17 & 0.33 & 3.0 \\ 
  &Dispersion (quad) & 0.2 & 0.04 & 0.54 & 3.0 \\ 
  \rowcolor[gray]{.9}& Influence  & 0.8 & 0.87 & 0.59 & 34.5 \\ 
\end{tabular}
\end{center}
\caption{Results of the power test for the simulation model with the data set reduced to N = 60 actors and 3 waves of data collection. The two parameters that relate to the hypotheses are highlighted gray.}
\label{tab:results1}
\end{table}

The key parameters (homophily and influence) are highlighted gray. Homophily has a power of 99.5\%, the influence effect a power of 34.5\%. 
Assuming that the simulated mathematical models are indeed a good representation of the real social processes, we could expect to find a significant influence effect in one out of three studies.  
This is not likely to be a sufficiently good expectation.
Note that some mean parameter estimates differ from the simulated values in Table~\ref{tab:results1} even though estimates of SAOMs in general are consistent with simulated values \citep{Block2017}. 
These deviations are explained by the fact that the simulation model was specified and run on a complete friendship network of 120 actors. Only after the simulation, a sub data set of 60 actors was extracted. This affects the estimates of all parameters that correlate with density-, clustering- and distance-related statistics. 
For example, propinquity matters less in this re-estimation that is based on just two communities. The density parameter, however, is more pronounced as it balances out the higher levels of network clustering and the smaller effect of the propinquity parameter. The parameter estimates are thus not unbiased in this example.
Still, the power of most of these network-related effects is high. 
The power of the attribute shape effects (linear and quadratic) is very low which is in line with our initial discussion that attribute-related effects are particularly prone to have a low statistical power. 

A comparison of the power of the five study designs is given in Table \ref{tab:power1}. The table now only focuses on the power estimates of the two key parameters homophily and influence that are related to the initial hypotheses. 
The columns express the study design decision about the network delineation which ranges from 120 actors (four communities) to 30 actors (one community). The rows show the varying number of data collection waves. The value in the table are again the percentages of models with significant results (at 5\% level) of the homophily (first value) and the social influence parameter (second value). These estimates of statistical power correspond to the right column in Table~\ref{tab:results1}.


\begin{table}[htb]
\renewcommand{\arraystretch}{1.3}
\begin{center}
\begin{tabular}{llcccccc}
  & & \multicolumn{6}{c}{\textbf{Community size}}\\
  & & \multicolumn{2}{c}{\textbf{N = 120}} & \multicolumn{2}{c}{\textbf{N = 60}} & \multicolumn{2}{c}{\textbf{N = 30}}\\
  & & Hom. & Inf. & Hom. & Inf.& Hom. & Inf.\\
\multirow{3}{*}{\textbf{Number of waves}} & \textbf{5 waves} & 100&97.5 & & & 97.5&28.5\\
  &\textbf{3 waves}  & & & 99.5&34.5 & &\\
  & \textbf{2 waves} & 99.5&34.5 & & & 34.5&10.0\\
\end{tabular}
\end{center}
\caption{Percentage of significant findings (in a 95\% confidence interval) of the homophily (first value) and the social influence parameter (second value) in five different cases in which sample size (number of local communities) and number of data collection waves vary. These power estimates are based on 200 simulations and re-estimations per parameter combination.}
\label{tab:power1}
\end{table}

In the minimal design with two waves and 30 actors the power of the influence effect is only 10\% and also the power of the homophily effect is low (34.5\%). The statistical power estimates of the three intermediate designs (120 actors and two waves, 60 actors and three waves, 30 actors and five waves) are similar to one another: The power of the homophily effect is high and ranges between 97.5 and 99.5\%, whereas the power of the influence effect is again low and ranges between 28.5 and 34.5\%. 
It is noteworthy that the information available for the estimation of nodal variables is similar in the three intermediate cases: 
One can loosely say that the information about nodal attributes doubles when the network size doubles (from 30 to 60 to 120) and also doubles when the number of periods doubles (from one period -- two waves -- to two periods to four periods). 
Thereby, the three intermediate designs exhibit the same information regarding \emph{nodal} attributes.
This equivalence cannot be upheld for the case of network variables because each additional actor in the network contributes multiple tie variables. 
Doubling the number of actors in a network will
more than double the number of observed tie variables while doubling the number of waves will only double the tie variables. The study design with two waves and N=120 will thus be likely to have more power for network effects than the design with N=30 and 5 waves.
Only the large study design with 120 actors and five waves of data collection has an excellent power of 100\% for the homophily and 97.5\% for the influence parameter. 

\subsection{Conclusions of the first power study}

Based on the five study design evaluations, researchers could now decide on how to conduct the longitudinal study on opinion and friendship network formation in the four local communities. 
The small scale study design (i.e., with a smaller N, and fewer waves), seems to be inadequately powered. If the influence hypothesis was of less interest, the most feasible of the three intermediate study designs could be chosen. Only the large study design promises good statistical power for the estimation of both homophily and influence effects.
To elaborate on the power of the influence effect, researchers might want to run further power studies with, for example, 120 actors and three waves, 60 actors and four waves, or 90 actors and three waves. This would mean going back to step~3 (define a set of potential study designs) of the six-step procedure.
These findings cannot be straightforwardly generalized to other contexts as they are sensitive to the characteristics of a specific research setting. However, they indicate that the statistical power of selection and influence processes can be strongly related to study design parameters such as network size and number of data collection waves.

\section{Research setting 2: Co-evolution of friendship and delinquency in 21 schools}
\label{sec:smoking}
\begin{figure}[htbp]
  \includegraphics[width = \textwidth]{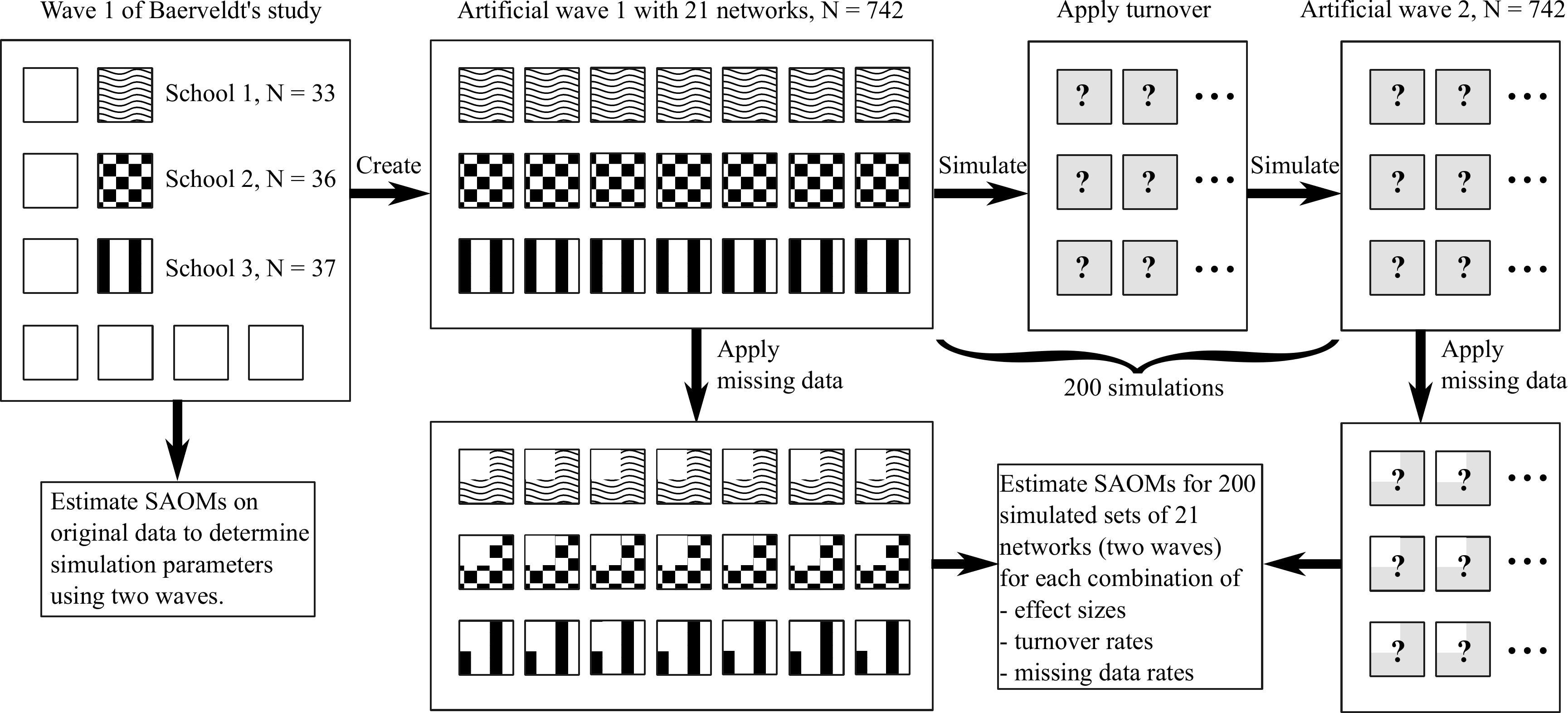}
\caption{The artificial school data set is based on three friendship networks (boxes with patterns; networks with sizes 33, 36 and 37 students) taken from the Baerveldt data. Seven additional networks were used for an estimation of parameters used in the simulation (indicated by empty boxes on the left). A second wave is simulated taking into account varying effect sizes, turnover rates, and missing data rates.}
\label{fig:data_study2}
\end{figure}

In the second research setting, we investigate how varying effect sizes, missing data and change in the composition of study participants may affect the power of selection and influence effects. 
We choose a setting that resembles a typical longitudinal network study in a population of schools and is inspired by the study of \citet{Baerveldt2008} on friendship selection and delinquency. We conduct a power study based on empirically observed friendship networks and delinquency attributes (measured on a five-point scale). 
The data preparation, simulation and estimation process is illustrated in Figure~\ref{fig:data_study2}.
First, we estimate a model that is similar to the one in the original study (using 10 networks and delinquency scores in a SAOM meta analysis). 
Second, we construct an artificial data set of 21 friendship networks that is based on three empirically observed networks. We use these 21 networks and the corresponding delinquency scores as the initial observation (wave 1).
Third, we simulate a second wave of data taking into account varying effect sizes, participant turnover (at half time between first and second wave), and missing data (applied after the simulation process and before the re-estimation). 
In total, 6,000 data sets are simulated. We use 30 combinations of effect sizes, participant turnover rates, and missing data rates. For each of these combinations, the set of 21 second wave networks and delinquency scores is simulated 200 times each (200~x~30~=~6,000).
Finally, SAOMs are estimated from the simulated data (using the SIENA multigroup option) and the power of the homophily and the influence effects is evaluated.

Compared to the first research setting, the number of participants is very high (N = 742 students, distributed over 21 schools). 
Data from three schools are replicated seven times each in order to construct the artificial sample.
Within the selected schools 33, 36, and 37 students are observed -- these are typical sizes of networks of age cohorts within the schools that \citet{Baerveldt2008} studied. 
This study focuses on how effect sizes, participant turnover (participants leaving and participants joining the population between waves) and missing data (participants not answering the questionnaire completely at random) affect the statistical power of the study design.
We again follow the six-step procedure proposed in section~\ref{sec:simulation_process}.

\subsection{Hypotheses and assumptions}
The key hypotheses are that both homophily and social influence processes regarding delinquency are prevalent within schools. In particular, we are interested in the effect of individuals selecting friends who are similar regarding the level of delinquency (homophily) as well as friendship network influence effects on student delinquency. As in research setting~1, we further assume the presence of a number of social network mechanisms (e.g., reciprocity, transitivity, gender homophily). Besides those we expect processes that result in participant turnover between data collection waves and missing data through non-participation. 
Unlike the first case study, which simulated data based on model parameters derived from the literature, Research Setting 2 uses results from an existing empirical data set
to inform parameter estimates. This relates to our advice to base initial assumptions on findings in related studies\footnote{We do not want to imply here that power studies should be performed using empirical results of the same study in an attempt to interpret the model parameters. We discuss the danger of post-hoc power studies in the conclusion section.}.
The rate of missing data, participant turnover, and homophily and influence effect sizes are assumed to be uncertain in the design phase of the study and so different values are compared to assess the sensitivity of the study design to these assumptions.

\subsection{Mathematical formulation}

We use the stochastic actor-oriented model to describe changes in the network structure and the individual delinquency variables. In the mathematical formulation we follow the empirical model of \citet[table 5, p.574]{Baerveldt2008} with some adaptations. For reasons of simplicity, some potentially relevant social mechanisms are omitted, for example, ethnic homophily. An effect capturing an interaction between reciprocity and transitivity \citep[Reciprocity in triads, see][]{Block2015} is added to the friendship model and a quadratic shape effect is included in the behavior change part of the model. Thereby, the model is closer to state-of-the-art SAOM specifications\footnote{The model of \citet{Baerveldt2008} is flawed because it omits the quadratic shape parameter that models dispersion of the behavioral variable. What they find as influence is essentially underdispersion that was not captured and hence appears as ``staying close to friends'' for a lack of closer effect in the model.}.
The complete specification of the SAOM is shown in table \ref{tab:specification2}. The parameters used for the simulation model were estimated on an empirical sample of 10 empirically observed school classes using a meta-analysis \citep{Snijders2003}. The focal parameters are highlighted gray. We test the power of parameters in two models: One in which we simulate effects that stem from a reanalysis of Baerveldt's data (``smaller'' effect sizes), and one in which we use slightly higher parameters (``larger'' effect sizes).

\begin{table}[htb]
\renewcommand{\arraystretch}{1.3}
\begin{center}
\begin{tabular}{rllr}
   &\textbf{Mechanism} & \textbf{SIENA effect name} &\textbf{Parameter}\\
\hline
   Friendship&Change & \code{rate} & ~4.3 \\
   &Density & \code{density} & -3.1\\
   &Reciprocity & \code{recip} & ~2.4\\
   &Transitivity & \code{transTrip} & 1.2 \\
   &Reciprocity in triads & \code{transRecTrip} & -0.8 \\
   &Homophily Sex & \code{sameX} & 0.6 \\
   \rowcolor[gray]{.9}&Homophily Delinquency & \code{simX} & smaller: 0.4 / larger: 0.6\\
   Delinquency & Change & \code{rate} & 1.3 \\
   &Center & \code{linear} & -0.2\\
   &Dispersion& \code{quad} & -0.2 \\
   \rowcolor[gray]{.9}&Influence & \code{avAlt} & smaller: 0.3 / larger: 0.4
\end{tabular}
\end{center}
\caption{Formal specification of the mathematical model in the second research settings. The focal mechanisms are highlighted gray.}
\label{tab:specification2}
\end{table}

This basic model is extended by two straightforward mechanisms. The first mechanism describes turnover of students after half of the data collection period, the second mechanism generates missing data that stems from completely random non-participation of some students in the two data waves (one empirical, one simulated). 

The turnover mechanism explains how students leave and join the sample. At half-time between the two data collection waves, a fixed number of students drops out of each school cohort (0, 1, or 3). At the same time, the same number of students joins the school so that the school size (ranging from 33 to 37 individuals) remains constant. The new students are network isolates in the moment they join the school and only then start forming friendship relations. 
The attributes of a new student are randomly chosen based on a frequency table of the attributes of all students (gender x delinquency) in the population at the time when the participant turnover occurs. 

The missing data mechanism relates to random non-participation in a survey wave. In both data collection waves a fixed number of students is selected from each of the seven school cohorts (0, 1, 3, 5, 7). Their network nominations and delinquency levels are treated as missing. The number of missing entries is the same in both data collection waves. The two random draws of missing individuals in the two waves are independent.

In this research setting we thus assume uncertainty about the levels of participant turnover (0, 1, 3 $\widehat{=}$  0\%, 2.8\%, 8.5\%), missing data (0, 1, 3, 5, 7 $\widehat{=}$ 0\%, 2.8\%, 8.5\%, 14.2\%, 19.8\%), and the effect size of homophily (simX in \{0.4, 0.6\}) and influence mechanisms (avAlt in \{0.3, 0.4\}). In total, there are 30 combinations of these three variables. 

\subsection{Potential study designs}

We do not consider different study designs. The statistical power of the mechanisms is tested for a study design that includes all 21 schools (N = 742 students), two waves of data collection, binary friendship nominations and a five-point delinquency scale. The space of alternative scenarios is therefore only defined by the rates of missing data, participant turnover rates, and the strength of selection and influence mechanisms.

\subsection{Simulation models}
The simulation models are based on the parameters in table \ref{tab:specification2} (one model with smaller and one with larger homophily and influence effect sizes) and all 15 combinations of participant turnover rates and missing rates (30 simulation models). 
Each simulation model is simulated 200 times with the RSiena software \citep{Ripley2016}. 
An R function was developed for the simulations that we conduct in this study. It combines RSiena-based simulations with the interfering processes of participant turnover and missing data.
The first wave of data is taken from the empirical data of \citet{Baerveldt2008}. A second data wave is simulated for each school separately. In total, 6,000 data sets are thereby generated (30 simulation models x 200 simulations) that include 21 networks and corresponding delinquency scores. 

The data have certain particularities.
The average degree is very low (1.4 ties, the maximum in-degree is 5) even though the school networks are relatively big (33, 36, and 37 individuals). The average level of delinquency is 1.8 on a scale that ranges from 0 to 4. The dispersion of delinquency values is low. Of 742 individuals only 56 (7.5\%) have a minimum score of 0, and 21 (2.8\%) have a maximum score of 4.
%

After conducting the simulations, we check the goodness of fit \citep{Ripley2016} of a small number of the simulated networks regarding degree distributions and triad census and compare those to the empirically observed second data wave. The simulated networks are found to be similar to the empirical networks by which we conclude that the simulation models are appropriate\footnote{The SIENA GOF function allows a systematic comparison between the simulated values and the empirically observed values (for each value of the respective statistic, e.g., degree distribution or triad census) and provides a p-value that relates to the null hypothesis that the real value were drawn from the distribution of simulated networks \citep{Lospinoso2012}. In neither of the tested cases this null hypothesis could be rejected.}.

\subsection{Estimation with RSiena}
Parameters are estimated for sets of 21 networks simultaneously with the RSiena software using the ``multigroup'' option \citep[section 11.1]{Ripley2016} for the analysis of multiple networks.
The re-estimation of one alternative scenario (consisting of 200 multigroup data sets) takes between one and eight hours on a computer with 24 cpus.
A computer cluster has been used for this step so that multiple SIENA re-estimations could be run in parallel.
The overall computation time was therefore also about eight hours.

\subsection{Evaluating the power}

\begin{figure}[htbp]
   \subfigure{
  \includegraphics[width = 0.5\textwidth, page = 1]{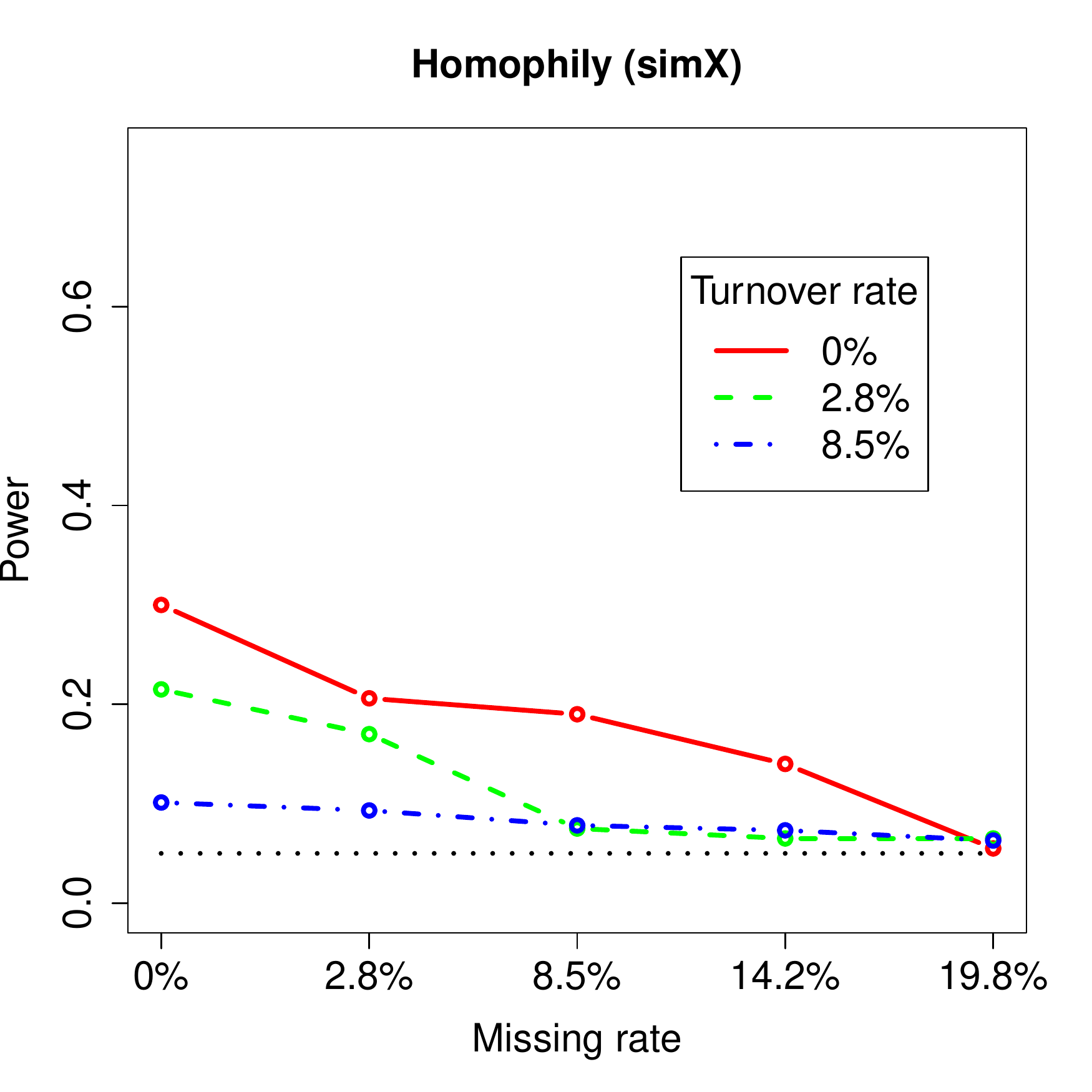}
  }
   \subfigure{
  \includegraphics[width = 0.5\textwidth, page = 2]{power_plots_avAlt03_simX04}
  }
\caption{Power of models with smaller homophily (simX = 0.4) and influence (avAlt = 0.3) parameters. Missing rates are indicated in the x axis, turnover rates are given by the three lines. 
The black dotted line indicates 
the chosen significance level (5\%).
}
\label{fig:power_2_1}
\end{figure}

\begin{figure}[htbp]
   \subfigure{
  \includegraphics[width = 0.5\textwidth, page = 1]{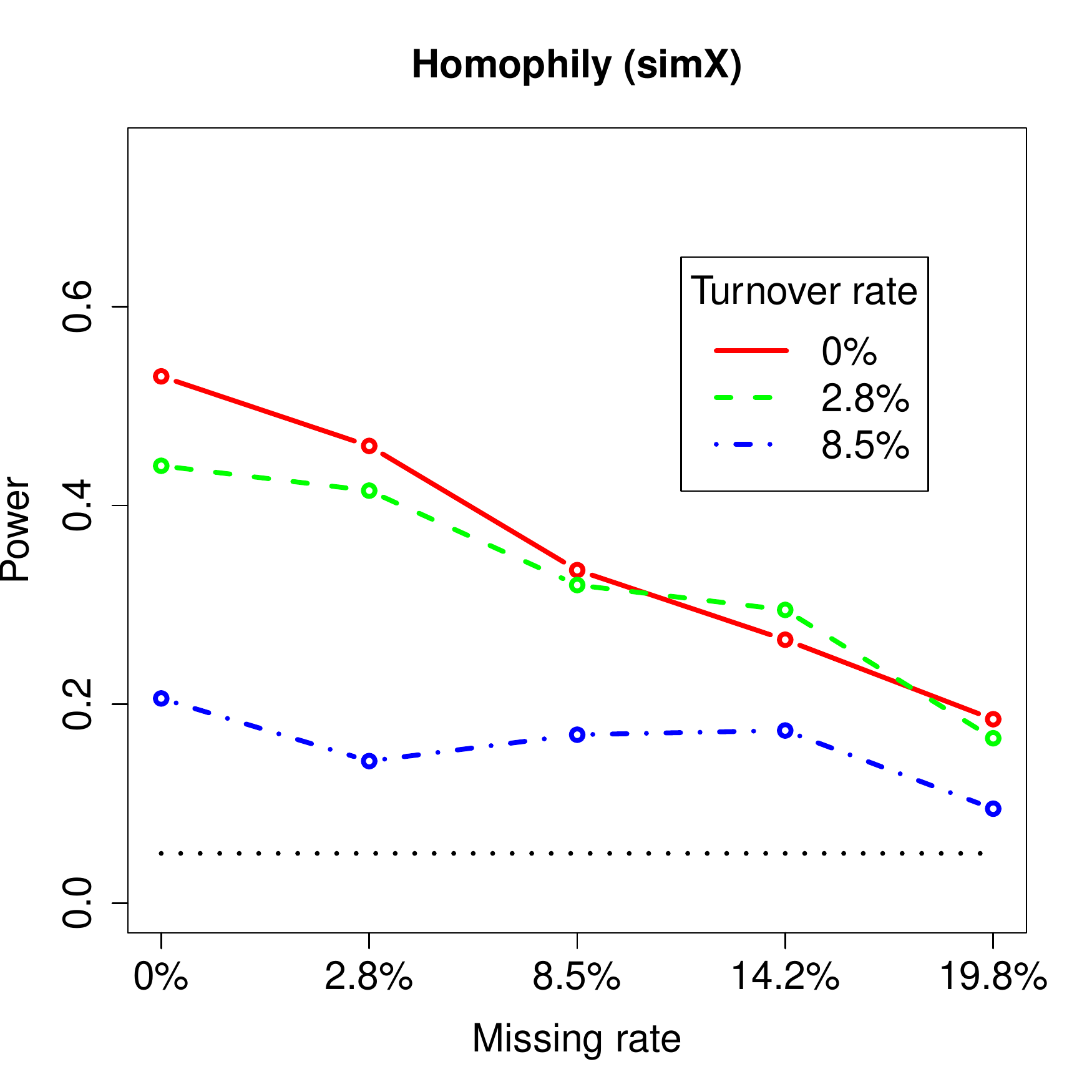}
  }
   \subfigure{
  \includegraphics[width = 0.5\textwidth, page = 2]{power_plots_avAlt04_simX06}
  }
\caption{Power of models with with larger homophily (simX = 0.6) and influence (avAlt = 0.4) parameters. Missing rates are indicated in the x axis, turnover rates are given by the three lines. 
The black dotted line indicates 
the chosen significance level (5\%).
}
\label{fig:power_2_2}
\end{figure}

The power estimates are given in Figures~\ref{fig:power_2_1} and~\ref{fig:power_2_2}. Figure~\ref{fig:power_2_1} shows the power estimates for the homophily and the influence parameter of model with smaller effect sizes (see Table~\ref{tab:specification2}), Figure~\ref{fig:power_2_2} those of the model with the larger effect sizes. Three lines indicate power of turnover rates of 0\%, 2.8\% and 8.5\%. The x-axis covers different missing rates. A dotted line at the 0.05 level 
indicates the chosen significance level that would be the expected power of unbiased estimates that have no information value at all (zero effects).
In both models, the power rates with no turnover and 2.8\% (low) turnover are somewhat similar and partly overlapping; a turnover of 2.8\% thus seems not to matter a lot. 
For example, the homophily parameter in the model with larger effect sizes (Figure~\ref{fig:power_2_2} on the left) has a power ranging from about 50\% (no missing data) to about 20\% (19.8\% missing data), irrespective of whether the turnover is zero or 2.8\% (the red and the green line).
However, there is a large drop in power with turnover rates of 8.5\% (the blue line). 
One problem that we encounter is that it is more difficult to achieve convergence of the estimation routine \citep[][sec.6]{Ripley2016} in case of models with an 8.5\% turnover rate and only two data waves. While close to 100\% of the models with zero and 2.8\% turnover converged, convergence could only be achieved in about 80\% of the high-turnover models.
The coverage rates under the null hypothesis of no effect are almost all sufficiently close to 0.95 (type I error close to 0.05) to conclude that under the null hypothesis the distribution of the parameter estimates is very close to a normal distribution with mean 0 and standard deviations equal to the reported standard error. The exception is the estimated social influence parameter (avAlt) in case of high-turnover (8.5\%) models, where the standard errors are inflated. With the small remaining sample size and the skewed dependent variable, this may be due to the occurrence of the so-called Donner-Hauck phenomenon \citep[][sec.8.1]{Hauck1977, Ripley2016} where the standard error is inflated and the Wald test should not be used for hypothesis testing. The very low rejection rates under the null are associated with lower power for the Wald test, if it would be used.
This explains why the power of the high-turnover models drops below the 5\% line in Figures~\ref{fig:power_2_1} and~\ref{fig:power_2_2}.
From a design point of view, the interpretation of the results is clear: with this amount of turnover for only two waves of data, it is impossible to have a satisfactory study of social influence.
In the following, we discuss results of the models in which the turnover rate was zero or 2.8\%.

In the models with weaker effects (Figure~\ref{fig:power_2_1}), the power of the homophily parameter and the influence parameter are rather low. The maximum power in a model without turnover and missings is 30\% (homophily, simX) and 38\% (social influence, avAlt). When the missing rates increase to 19.8\% the power of the homophily parameter drops to the random expectation of a null effect when a significance criterion of $\alpha$ = 0.05 is chosen (5\% power). The power of the influence effect remains only slightly higher.

The models with larger homophily and influence effect sizes (Figure~\ref{fig:power_2_2}) start off from higher power values. In case of no missing and no turnover the power of the larger homophily effect is 53\%, the power of the larger influence effect is 70\%. A turnover rate of 2.8\% seems not to affect the power estimates a lot. 
In a model with 19.8\% missing rates, the statistical power drops to 19\% and 22\% for homophily and social influence respectively.

\subsection{Conclusions of the second power study}

The second case study illustrates the potentially crucial effect of turnover and missing data on the power of a longitudinal study design. 
In some of the scenarios, the chances of detecting a real effect is not much larger than the chances of identifying a significant effect when the true effect is null: this is clearly nowhere near an acceptable or useful study design.
Missing data of 19.8\% (the highest simulated value) reduces the power greatly. The power of the influence parameter in the model with smaller effect sizes, for example, dropped from 37.5\% to 7.5\%. The latter is close to the 
type I error.
Advanced missing data imputation strategies might be able to reduce the effect of missing data on power \citep{Krause2018}.
Turnover also has a negative effect on power. We further observed an inflation of standard errors, probably due to the so-called Donner-Hauck phenomen. It turned out that with just two waves of data and a turnover rate of 8.5\% the statistical power was unsatisfying in all simulation models.

A notable observation is further that the power of the homophily parameter is generally lower than the power of the influence parameter. This seems counterintuitive given our initial discussion that homophily inference is based on $N \cdot \overline{k}$ observations while influence effects are estimated based on $N$ observations per wave. In this example, however, we use data with specific particularities that probably strongly affect the power of the study. First, the network is very sparse. Initially, only 1.4 friendship nominations exist ($\overline{k}=1.4$) which reduced the typical advantage of more information on testing dyadic hypotheses. At the same time, we estimate a higher number of effects in the network change sub-model (seven as compared to four in the behavior change sub-model) which might be related to a lower expected power. Second, the dispersion of the delinquency variable is very low; only 7.5\% and 2.1\% of individuals were in the lowest and highest category of the five-point scale in the first data wave. The homophily and the influence parameter are estimated based on cross-lagged statistics \citep{Steglich2010} that do not carry a lot of information when the variable dispersion is low and only few ties are observed. Researchers facing this problem might for example want to consider using a more fine-grained delinquency scale that generates a higher dispersion. This might improve the power of the homophily parameter in particular. As an improved estimation strategy it should be considered to use a maximum likelihood routine \citep{Snijders2010a} as it uses information more efficiently which may lead to an increased power. Using maximum likelihood estimations in the re-estimation of simulated models (step five of the six-step routine) is possible in general but will take much more time. 

\section{Discussion and conclusions}
\label{sec:conclusions}

In this paper, we presented a procedure for performing power analyses in longitudinal social network studies. In particular, we discussed study designs that aim at investigating social selection and influence mechanisms with stochastic actor-oriented models (SAOMs). 
About 130 empirical studies of that type have been published in the recent years \citep{Snijders2016}. Those studies report mixed findings about homophily and social influence processes which we argued might be related to power issues.
The six-step procedure that we presented in this paper can be seen as a tool for the investigation and comparison of statistical power of longitudinal social network study designs. We demonstrated its utility in two extensive research settings that focused on the effect of network size, number of data collection waves, effect sizes, missing data, and participant turnover on statistical power. 

The two research settings that we presented did not aim at providing practical rules of thumb because we are not yet at the point where general conclusions and design recommendations can be formulated. Nevertheless, they made clear that network delineation, number of data collection waves, effect sizes, missing data and participant turnover may strongly affect the power of longitudinal selection and influence studies. In research setting~1 (section~\ref{sec:local_communities}), we specified a mathematical model of selection and influence with pronounced effect sizes. A simulated small-scale study design with 30 individuals and two waves of data collection was found to be inappropriate for empirically testing either of the two effects. 
A study design with five waves of data and 120 individuals provided excellent power for both the homophily and the influence effect. 
In research setting~2 (section~\ref{sec:smoking}), we specified a similar mathematical model for selection and influence dynamics among 742 students distributed over 21 schools. 
The simulated effect sizes in this study were smaller, we only simulated two data waves, and the initial data carried a lot less information.
Given those study characteristics, we found that a missing data rate of 20\% would strongly reduce the power of homophily and influence parameters. In a simulation model with low effect sizes, the power was 
not meaningfully larger than the level of significance.
A turnover rate of 8.5\% also had a strongly negative effect on statistical power. A practical issue that arose in models with high participant turnover is that it is harder to achieve convergence in the estimation routine.
Missing data and participant turnover rates in that magnitude are not uncommon. This underlines the importance of social network data collections that aim at high participation rates and panel stability over time.

The two empirical settings provide some intuition about issues that researchers should be concerned about, however, the  quantitative results should not be generalized. We could indeed show that in these cases the power estimates are highly affected by variations in a number of study design and social mechanism parameters. Those parameters jointly affect the power.
For example, we discussed that the distribution of variables and the network structure affected the power in study designs in which we also modeled high participant turnover.
We also showed that assumptions about parameter values matter. When researchers face uncertainty, it is advisable not to define just one simulation model, but several models with varying parameters as we illustrated in the second research setting.

A question that is likely to arise from this work is whether the procedure may be used to investigate
if insignificant effects in an empirical study result from a lack of statistical power. However, it is common sense among statisticians that post-hoc power studies are \emph{irrelevant in the interpretation of empirical results} \citep{Cox1958, Goodman1994, Senn2002, Lenth2007}. 
Estimating the power of a study design as a result of not finding significant evidence for a hypothesis may lead to the dangerous conclusion that evidence for a (non-significant) social mechanism may just not have been found because of a lack of power.
Yet, the level of confidence about an estimate is already captured by the estimated standard errors or confidence intervals. 

Post-hoc power studies should thus never be used in the interpretation of parameters. However, they may motivate future research in case they suggest that certain adaptations may indeed improve the power of a study design. \citet{Gelman2013} propose that post-hoc ``design analyses'' may generally be useful when assumptions about social mechanisms stem from prior expectations or prior empirical findings but not from the empirical estimates.
They argue that design analyses that are ``based on an effect size that is determined from literature review or other information external to the data at hand can be helpful in reflecting on the results'' \citep[][p.2]{Gelman2013} irrespective of whether the findings are significant or not.


The six-step procedure proposed can provide new guidelines for the design of longitudinal social network studies. We hope that it will inspire systematic investigation of longitudinal study designs on various dimensions. In our examples, we showed that network size, duration of a study, effect sizes, missing data and participant turnover mattered for statistical power. Other directions are to be explored in the future: How do, for example, assumptions about measurement scales, systematic types of missing data, varying assumptions about interfering social mechanisms, alternative influence mechanisms, measurement errors, and varying time intervals affect the power of a study design?
Many of these topics are of critical importance for empirical research and should thus be explored in varying contexts in the future. The six-step procedure that we presented in this article is an adequate tool to do develop a deeper understanding of statistical power in longitudinal network studies.

\newpage
\renewcommand{\makeenmark}{\theenmark~}
\renewcommand{\notesname}{ENDNOTES}
\theendnotes


\newpage
\renewcommand{\refname}{REFERENCES}

\bibliographystyle{ajs} 
\bibliography{power_paper}

\end{document}